\documentclass[conference]{IEEEtran}
\IEEEoverridecommandlockouts
\usepackage{cite}
\usepackage{amsmath,amssymb,amsfonts}
\usepackage{algorithmic}
\usepackage{graphicx}
\usepackage{textcomp}
\usepackage{balance}
\usepackage{hyperref}
\usepackage[nolist]{acronym}
\usepackage{xcolor}
\usepackage{caption}
\usepackage{subcaption}
\usepackage{soul}
\usepackage[inline]{enumitem}
\usepackage{cleveref}
\usepackage{tikz}
\crefname{section}{Sec.}{Secs.}
\crefname{figure}{Fig.}{Figs.}
\crefname{equation}{Eq.}{Eqs.}

\newcommand{\MARVEL}{\href{https://www.marvel-project.eu/}{H2020 MARVEL project}}

\begin{document}
\bstctlcite{IEEEexample:BSTcontrol}
\begin{acronym}
  \acro{3GPP}{Third Generation Partnership Project}
  \acro{5G-PPP}{5G Public Private Partnership}
  \acro{AA}{Authentication and Authorization}
  \acro{ADAS}{Advanced Driver-Assistance Systems}
  \acro{AI}{Artificial Intelligence}
  \acro{API}{Application Programming Interface}
  \acro{AP}{Access Point}
  \acro{AR}{Augmented Reality}
  \acro{AV}{Audio-Video}
  \acro{BGP}{Border Gateway Protocol}
  \acro{BSP}{Bulk Synchronous Parallel}
  \acro{BS}{Base Station}
  \acro{CDF}{Cumulative Distribution Function}
  \acro{CFS}{Customer Facing Service}
  \acro{CPU}{Central Processing Unit}
  \acro{DHT}{Distributed Hash Table}
  \acro{DMP}{Data Management Platform}
  \acro{DNS}{Domain Name System}
  \acro{E2F2C}{Edge to Fog to Cloud}
  \acro{ETSI}{European Telecommunications Standards Institute}
  \acro{FCFS}{First Come First Serve}
  \acro{FSM}{Finite State Machine}
  \acro{FaaS}{Function as a Service}
  \acro{GPU}{Graphics Processing Unit}
  \acro{HMI}{Human Machine Interface}
  \acro{HTML}{HyperText Markup Language}
  \acro{HTTP}{Hyper-Text Transfer Protocol}
  \acro{ICN}{Information-Centric Networking}
  \acro{IETF}{Internet Engineering Task Force}
  \acro{IIoT}{Industrial Internet of Things}
  \acro{IPP}{Interrupted Poisson Process}
  \acro{IP}{Internet Protocol}
  \acro{ISG}{Industry Specification Group}
  \acro{ITS}{Intelligent Transportation System}
  \acro{ITU}{International Telecommunication Union}
  \acro{IT}{Information Technology}
  \acro{IaaS}{Infrastructure as a Service}
  \acro{IoT}{Internet of Things}
  \acro{JSON}{JavaScript Object Notation}
  \acro{LCM}{Life Cycle Management}
  \acro{LL}{Link Layer}
  \acro{LTE}{Long Term Evolution}
  \acro{MAC}{Medium Access Layer}
  \acro{MBWA}{Mobile Broadband Wireless Access}
  \acro{MCC}{Mobile Cloud Computing}
  \acro{MEC}{Multi-access Edge Computing}
  \acro{MEH}{Mobile Edge Host}
  \acro{MEPM}{Mobile Edge Platform Manager}
  \acro{MEP}{Mobile Edge Platform}
  \acro{ME}{Mobile Edge}
  \acro{ML}{Machine Learning}
  \acro{MNO}{Mobile Network Operator}
  \acro{NAT}{Network Address Translation}
  \acro{NFV}{Network Function Virtualization}
  \acro{NFaaS}{Named Function as a Service}
  \acro{OSPF}{Open Shortest Path First}
  \acro{OSS}{Operations Support System}
  \acro{OS}{Operating System}
  \acro{OWC}{OpenWhisk Controller}
  \acro{P2P}{Peer-to-Peer}
  \acro{PMF}{Probability Mass Function}
  \acro{PU}{Processing Unit}
  \acro{PaaS}{Platform as a Service}
  \acro{PoA}{Point of Attachment}
  \acro{QoE}{Quality of Experience}
  \acro{QoS}{Quality of Service}
  \acro{RPC}{Remote Procedure Call}
  \acro{RR}{Round Robin}
  \acro{RSU}{Road Side Unit}
  \acro{SAN}{Storage Area Network}
  \acro{SBC}{Single-Board Computer}
  \acro{SDN}{Software Defined Networking}
  \acro{SDK}{Software Development Kit}
  \acro{SLA}{Service Level Agreement}
  \acro{SMP}{Symmetric Multiprocessing}
  \acro{SRPT}{Shortest Remaining Processing Time}
  \acro{STL}{Standard Template Library}
  \acro{SaaS}{Software as a Service}
  \acro{TCP}{Transmission Control Protocol}
  \acro{TSN}{Time-Sensitive Networking}
  \acro{UDP}{User Datagram Protocol}
  \acro{UE}{User Equipment}
  \acro{URI}{Uniform Resource Identifier}
  \acro{URL}{Uniform Resource Locator}
  \acro{UT}{User Terminal}
  \acro{VANET}{Vehicular Ad-hoc Network}
  \acro{VIM}{Virtual Infrastructure Manager}
  \acro{VM}{Virtual Machine}
  \acro{VNF}{Virtual Network Function}
  \acro{VR}{Virtual Reality}
  \acro{WLAN}{Wireless Local Area Network}
  \acro{WMN}{Wireless Mesh Network}
  \acro{WRR}{Weighted Round Robin}
  \acro{YAML}{YAML Ain't Markup Language}
\end{acronym}

\title{Design Guidelines for Apache Kafka Driven Data Management and Distribution in Smart Cities
\thanks{This work was funded by the European Union's Horizon 2020 research and innovation programme MARVEL under grant agreement No 957337. This publication reflects the authors views only. The European Commission is not responsible for any use that may be made of the information it contains.}
}

\author{\IEEEauthorblockN{Theofanis P. Raptis\IEEEauthorrefmark{1},
Claudio Cicconetti\IEEEauthorrefmark{1}, 
Manolis Falelakis\IEEEauthorrefmark{2}, Tassos Kanellos\IEEEauthorrefmark{3},
Tomás Pariente Lobo\IEEEauthorrefmark{4}
\IEEEauthorblockA{\IEEEauthorrefmark{1}Institute of Informatics and Telematics, National Research Council, Pisa, Italy. Email: \{name.surname\}@iit.cnr.it}
\IEEEauthorblockA{\IEEEauthorrefmark{2}Netcompany-Intrasoft, Athens, Greece. Email: manolis.falelakis@netcompany-intrasoft.com}
\IEEEauthorblockA{\IEEEauthorrefmark{3}ITML, Athens, Greece. Email: tkanellos@itml.gr}
\IEEEauthorblockA{\IEEEauthorrefmark{4}Atos Spain, Madrid, Spain. Email: tomas.parientelobo@atos.net}
}
}

\maketitle
\begin{tikzpicture}[remember picture,overlay]
\node[anchor=south,yshift=10pt] at (current page.south) {\fbox{\parbox{\dimexpr\textwidth-\fboxsep-\fboxrule\relax}{
  \footnotesize{
    This work has been submitted to the IEEE for possible publication. Copyright may be transferred without notice, after which this version may no longer be accessible.
  }
}}};
\end{tikzpicture}

\begin{abstract}
Smart city management is going through a remarkable transition, in terms of quality and diversity of services provided to the end-users. The stakeholders that deliver pervasive applications are now able to address fundamental challenges in the big data value chain, from data acquisition, data analysis and processing, data storage and curation, and data visualisation in real scenarios. Industry 4.0 is pushing this trend forward, demanding for servitization of products and data, also for the smart cities sector where humans, sensors and devices are operating in strict collaboration. The data produced by the ubiquitous devices must be processed quickly to allow the implementation of reactive services such as situational awareness, video surveillance and geo-localization, while always ensuring the safety and privacy of involved citizens. This paper proposes a modular architecture to (i) leverage innovative technologies for data acquisition, management and distribution (such as Apache Kafka and Apache NiFi), (ii) develop a multi-layer engineering solution for revealing valuable and hidden societal knowledge in smart cities environment, and (iii) tackle the main issues in tasks involving complex data flows and provide general guidelines to solve them. We derived some guidelines from an experimental setting performed together with leading industrial technical departments to accomplish an efficient system for monitoring and servitization of smart city assets, with a scalable platform that confirms its usefulness in numerous smart city use cases with different needs.
\end{abstract}


\section{Introduction}\label{sec:introduction}

During the last 10--15 years, there has been an explosion of enabling technologies for the realization of the \ac{IoT}, including sensors, actuators, embedded devices with computation capabilities, software platforms, and communication protocols~\cite{7123563,7879243}.
This phenomenon was driven initially by the huge potential foreseen in the automation and digitization of industrial applications~\cite{6714496} and personal health systems~\cite{7113786}, but it benefited many other segments through spillover effects.
One of the most important outlets of the growing \ac{IoT} ecosystem has been the smart city market~\cite{6740844}, which was ready to incorporate new technologies to supply citizens, as well as city councils, with new services or more efficient realizations of existing ones.

\begin{figure*}[t!]
    \centering
    \includegraphics{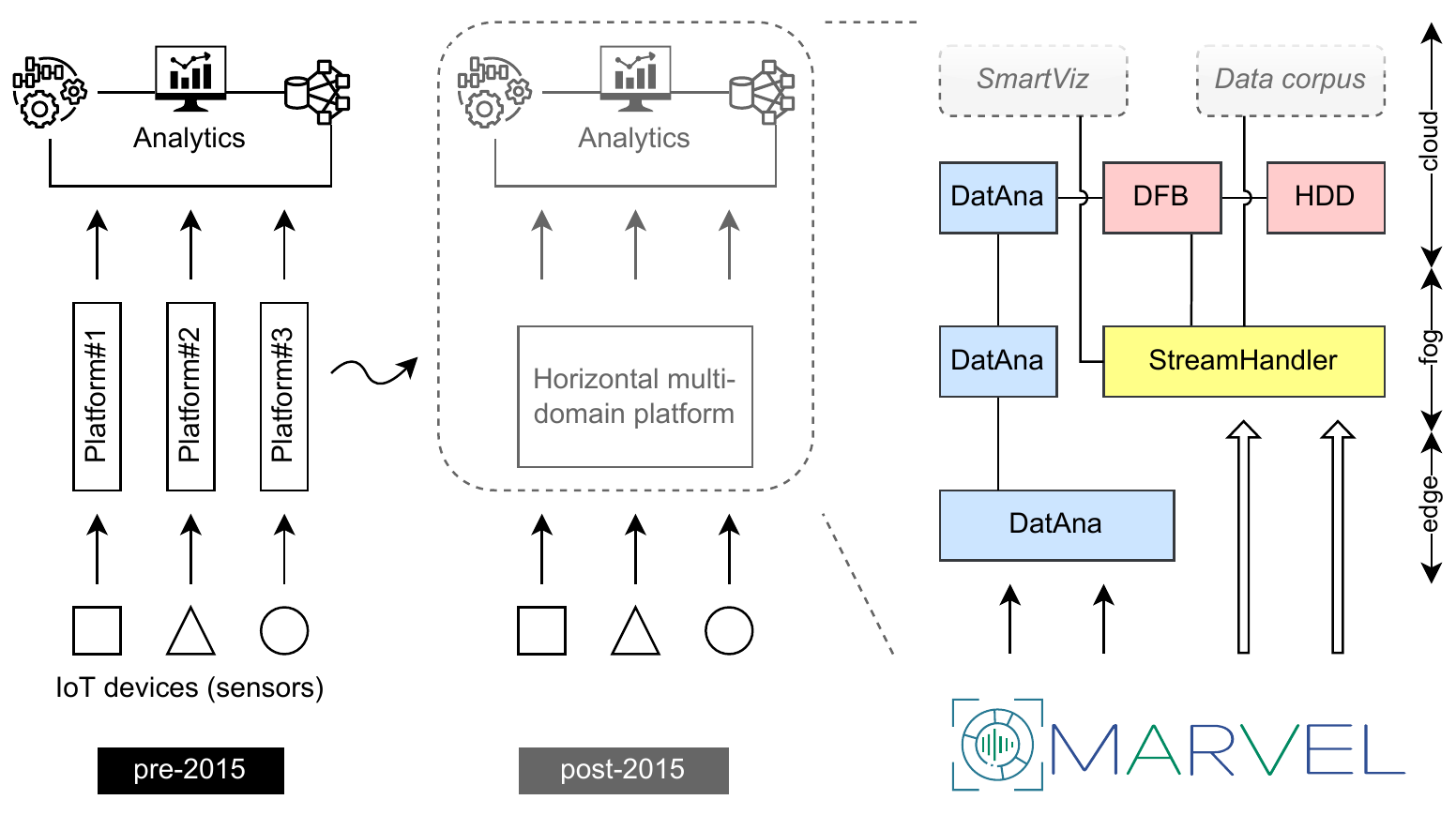}
    \caption{Transition of smart city data management and distribution systems from vertical silos to horizontal (cloud-based) platforms and positioning of the \MARVEL's proposition for multi-layer \ac{E2F2C} environments.}
    \label{fig:high-level}
\end{figure*}

In the early developments of smart cities, each service relied on its own devices that could operate only with a dedicated proprietary platform in a vertical manner.
Typically, the platforms offered \acp{API} for the consolidation of data across multiple services in the cloud, e.g., for integrated user dashboards or big data analysis of historical data.
Indeed, many studies have focused on supporting semantic interoperability of data only \textit{after} they have been safely stored in a common repository (currently referred to as ``data lake'')~\cite{6605518,7226706}.
However, such a compartmentalized structure had limitations, especially in terms of redundant deployed resources and inefficient management.
Therefore, the community has moved towards a horizontal approach, where a common platform is able to communicate with all kinds of devices; most often sensors in smart city applications \cite{9039732}.

This evolution is illustrated in \cref{fig:high-level}, which also shows the high-level architecture of the \ac{DMP} defined  the \href{https://www.marvel-project.eu/}{H2020 MARVEL project}, which aspires to define a comprehensive solution for multi-modal real-time analytics applications.
Such applications derive from the analysis of the requirements and expectations in several use cases of practical interest and high impact of the quality of life of citizens, and they will be validated in three small-scale field trials across Europe, i.e., in Malta, Serbia, and Italy~\cite{9593258}.
In the project, we exploit the recent trend of breaking down the computation elements of the system into three layers in a hierarchy~\cite{Ahmed2017,Porambage2018}: the \textit{edge} layer is closest to the sensing and embedded computation devices, but it consists of devices with modest capabilities in terms of computation, connectivity, and storage; the \textit{fog} layer has more powerful capabilities and we treat it like a small private cloud, which is however under the control of the end user; and, finally, the \textit{cloud} layer is hosted on public remote data centers, which have virtually infinite capacity but incur a high latency and usage costs.

The \MARVEL\ covers all aspects, from the development of new sensors, e.g., directional microphone arrays, to the efficient training of \ac{AI}/\ac{ML} models on devices with limited capabilities, to the ethics of data collection and analysis.
However, in this paper we focus only on the \ac{DMP}, which is the core of the project's software architecture and consists of the following components:

\begin{itemize}[parsep=0em,leftmargin=*,label={--}]%
    \item \textbf{DatAna}, for the processing and transmission of structured data produced by AI components to all the layers in the inference pipeline;
    \item \textbf{DFB}, which is in charge of managing heterogeneous data across multiple components in the cloud; 
    \item \textbf{StreamHandler}, which processes, stores and delivers real-time AV data at the fog layer toward the processing servers and the Data corpus;
    \item \textbf{HDD}, which can optimize the data management based on the available resources and current workload;
\end{itemize}

Those core components work in close connection to the \textbf{Data corpus}, which is the repository of the data collected from the sensors, mainly consisting of microphones and cameras, for visualization and augmentation, and the \textbf{SmartViz}, which provides the \ac{HMI} for visualization and analysis.
    
The rest of the paper is structured as follows.
First, we introduce the background and foundations of the \MARVEL, which are needed to understand the concepts and terminology used in the rest of the paper, in \cref{sec:background}.
This section also includes an overview of the essential state of the art on data/resource management in edge/fog systems.
In \cref{sec:dmp} we then illustrate all the main components of the \MARVEL's \ac{DMP}.
Handling \ac{AV} data was found to be challenging in particular, hence it is discussed separately in \cref{sec:avdata}.
Preliminary results obtained during the mid-project integration tests are reported in \cref{sec:eval}, while \cref{sec::guidelines} presents some useful design guidelines which directly reflect our experience.

\section{Background}\label{sec:background}

In this section we provide an overview of the aspects of the \MARVEL\ that are relevant to the design and implementation of the \ac{DMP}, which is the main subject of this work.

\begin{figure}[t!]
    \centering
    \includegraphics[width=\columnwidth]{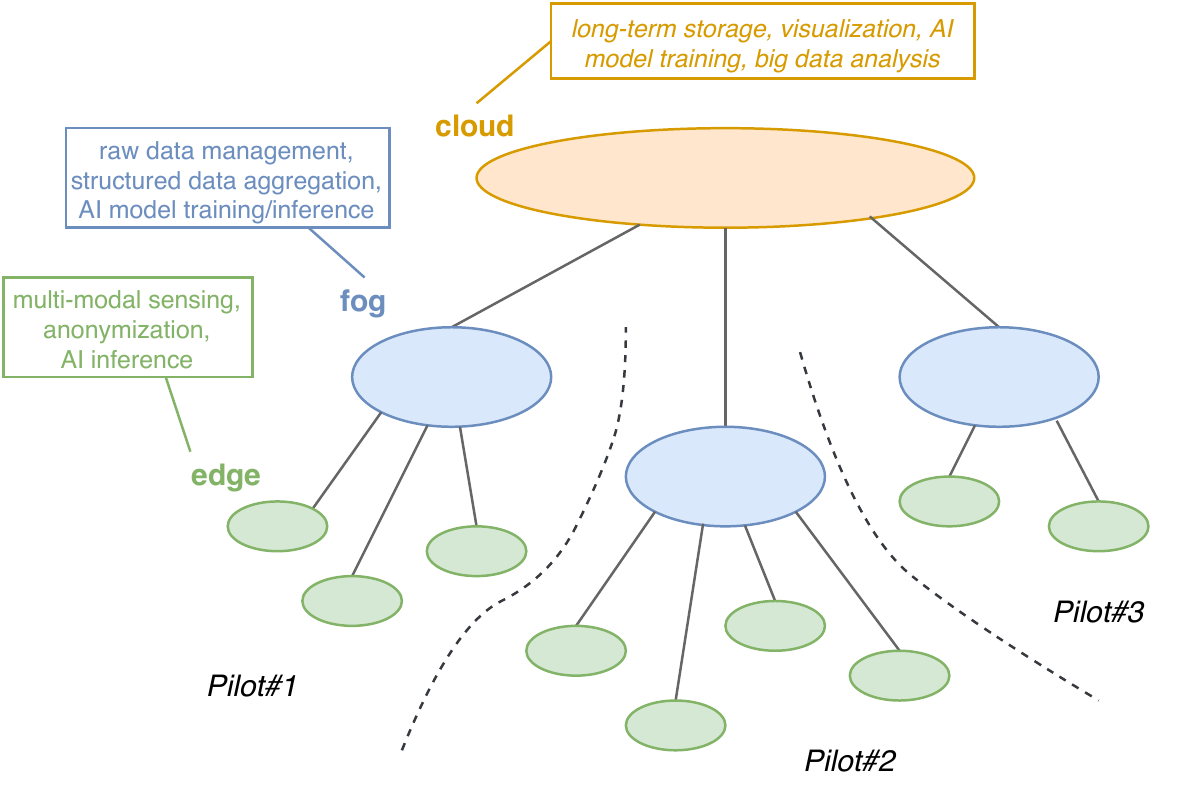}
    \caption{\ac{E2F2C} view of the \MARVEL's pilots.}
    \label{fig:layered-view}
\end{figure}

A layered view of the physical deployment of the project's pilots is illustrated in \cref{fig:layered-view}.
The lowest layer is the \textit{edge}, which contains the sensors and embedded devices to perform on-site operations.
We have two types of sensors: microphone arrays, producing audio streams, and cameras, producing AV streams.
All the sensors are expected to operate continuously during the service lifetime.
The embedded devices, i.e., Raspberry Pis and NVIDIA Jetson boards, perform operations directly at the edge, which include anonymization of the AV streams and basic inference operations.
The intermediate layer is called the \textit{fog}\footnote{We not that in the scientific literature and in the market press the terms ``edge'' and ``fog'' do not have universally accepted meanings. Sometimes they are even used interchangeably.}, which includes more powerful computation resources that are shared by multiple edge sites, e.g., workstations and rack-mountable servers with \acp{GPU}, which are are suitable for not only managing the different data streams but also performing more advanced inference tasks, as well as training of \ac{AI} models.
Finally, the \textit{cloud} is hosted in an infrastructure provided by a project partner and it is common for all the pilots.
The cloud provides long-term storage of data and all the services for \ac{HMI}, i.e., visualization and real-time (on-demand) analysis.

In \cref{fig:layered-view} we distinguish between raw data vs.\ structured data:

\begin{itemize}[parsep=0em,leftmargin=*,label={--}]%
    \item The \textbf{raw data} are the AV streams generated by the sensors, irrespective of whether they have been anonymized or not.
    Their formats and characteristics are heterogeneous because they depend on the physical devices installed (e.g., may use different codec or sample AV at different rates).
    In any case, the throughput is generally high, especially for video, which requires carefully provisioned bandwidth, long-term storage, and configuration of the data distribution services.
    \item The \textbf{structured data}, instead, are data generated by the analytics applications, i.e., the inference components.
    In the project, have different data models for each AI component, but all of them have been based on the fully  Smart-Data-Models (SDM)-compliant\footnote{ https://smartdatamodels.org/} data models that DatAna is producing.
    Furthermore, their throughput is generally much lower than than of raw data.
\end{itemize}

\begin{figure*}[t!]
    \centering
    \includegraphics{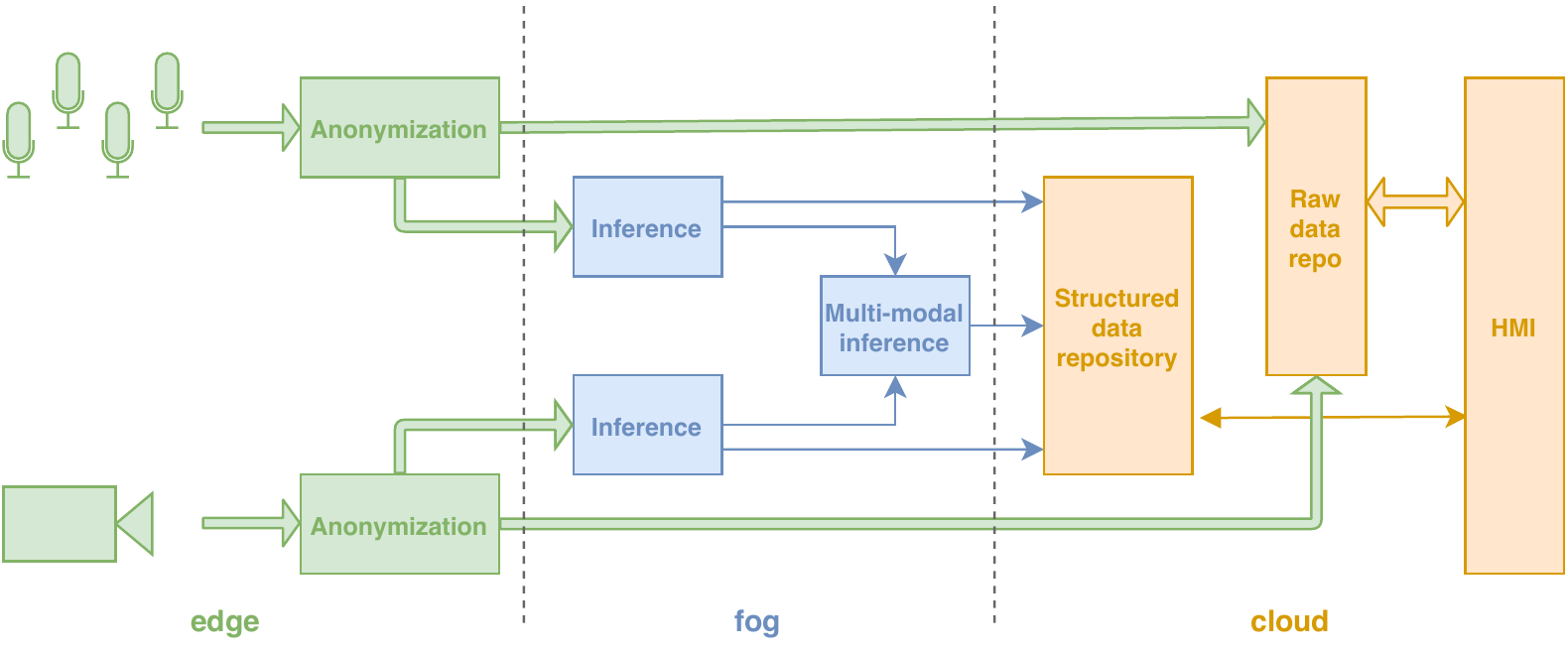}
    \caption{Schematic of a typical multi-modal, i.e., audio + video, real-time analytics application in \MARVEL. The mapping of components to the edge/fog/cloud layers is only indicative: in real applications this may depends on physical, environmental, and administrative constraints.}
    \label{fig:example-app}
\end{figure*}

A schematic of a typical multi-modal application in the \MARVEL\ is illustrated in \cref{fig:example-app}.
We have used ``thick'' arrows to represent the raw data flows, while ``thin'' arrows refer to structured data flows, to reflect in a pictorial manner their different bandwidth requirements.
The diagram is intended to provide the reader with a sketch of the components that are involved in the deployment of a service, which can differ in practice depending on the specific use case and pilot.
The services we support in the project include surveillance applications (e.g., detecting anomalies in public spaces), emotion recognition (with aerial images taken by drones), road safety (e.g., monitoring junctions with mixed cars and bike traffic).

\subsection{Related work}

The topic of efficient allocation of resources in \ac{E2F2C} has been extensively studied in the scientific literature (e.g., \cite{Wang2019a}).
The vast majority of the works propose highly simplified mathematical models, which are then solved to maximize a given objective function, but they do not linger on the practical implementation of their solution using industry-grade tools.
On the other hand, in the \MARVEL\ our goal was to realize a solution based on widely-used, reliable, and open source software.
Furthermore, some works have focused on the specific aspect of efficient data distribution.
In~\cite{Becker2021}, the authors have proposed \ac{P2P} as a means to distribute data in a robust and decentralized manner among agents at the edge.
Such an approach only partially covers the needs of the \MARVEL, since AV real-time streams cannot be stored efficiently in a \ac{P2P} overlay.
An in-network storage management is put forward in~\cite{Nicolaescu2021}, to place both raw and structured data on distributed resources using the Google File System, which is however proprietary.
Another perspective is taken in~\cite{Gupta2018}, which builds on top of the distributed database \href{https://cassandra.apache.org/}{Apache Cassandra} by adding geolocalization tags that help distributing the load across edge/fog nodes; we have not considered this feature since, at least in the project pilots, the edge/fog nodes are all deployed in close proximity.
Finally, in~\cite{9193994}, the authors distribute the execution of functions following a \ac{FaaS} approach, which is adopted in production only in cloud systems, but has shown some potential also for edge computing~\cite{Aslanpour2021}, even though further research may be needed. 
The solutions mentioned can be gradually incorporated in the \ac{DMP} at a later stage of the project, which is still ongoing with expected completion date at the end of 2023.

More specifically, Apache Kafka has also been widely used in the streaming applications domain. In \cite{10.14778/2824032.2824063}, the authors replicate Apache Kafka logs for various distributed data-driven systems at LinkedIn, including source-of-truth data storage and stream processing. In \cite{9016008}, the authors design a distributed cluster processing model based on Apache Kafka data queues, to optimize the inbound efficiency of seismic waveform data. In \cite{9378217}, the authors extend Apache Kafka by building an in-memory distributed complex event recognition engine built on top of Apache Kafka streams. In \cite{10.1007/978-3-030-84811-8_2}, the authors design a simulation platform enabling evaluations of future mobility scenarios, based on an Apache Kafka architecture. In \cite{10.1145/3148055.3148068}, the authors break the streaming pipeline into two distinct phases and evaluate percentile latencies for two different networks, namely 40GbE and InfiniBand EDR (100Gbps), to determine if a typical streaming application is network intensive enough to benefit from a faster interconnect. Moreover, they explore whether the volume of input data stream has any effect on the latency characteristics of the streaming pipeline, and if so, how does it compare for different stages in the streaming pipeline and different network interconnections. 
In \cite{10.1145/3445945.3445949}, the authors propose a distributed framework for the application of stream processing on heterogeneous environmental data, which addresses the challenges of data heterogeneity from heterogeneous systems and offers real-time processing of huge environmental datasets through a publish/subscribe method via a unified data pipeline with the application of Apache Kafka for real-time analytics. In \cite{10.14778/3137765.3137771}, the authors find that filtering on large datasets is best done in a common upstream point instead of being pushed to, and repeated, in downstream components. To demonstrate the advantages of such an approach, they modify Apache Kafka to perform limited native data transformation and filtering, relieving the downstream Spark application from doing this. Their approach outperforms four prevalent analytics pipeline architectures with negligible overhead compared to standard Kafka.
In the next sections, we illustrate the high-level design and current status of development of \MARVEL's \ac{DMP}.

\section{DMP Software Architecture Design}\label{sec:dmp}

The main components of the \ac{DMP} are introduced briefly in \cref{sec:introduction} (see also \cref{fig:high-level}).
Below we provided an overview, followed by component details in dedicated sub-sections.

\textbf{DatAna} is responsible for collecting the inference results from all AI components from all layers through its instances residing at each layer, transforming them into SDM-compliant counterparts and then transferring them to higher layers in the E2F2C continuum.
The DatAna cloud layer thus aggregates all transformed inference results and relays them to the DFB, which also resides at the cloud.
\textbf{DFB} persistently stores all SDM-compliant inference results it receives, but also makes them available in real time to SmartViz and Data Corpus.
DFB also exposes a REST API to SmartViz to allow it to access all archived inference results in its \href{https://www.elastic.co/}{ElasticSearch} database.
In addition, the DFB receives user-generated inference result verification messages from SmartViz and use this information to update the corresponding inference results stored in its database.
\textbf{HDD} interacts exclusively with the DFB to receive information on current Kafka topic partitioning and associated performance metrics and to send updated, optimised Kafka topic partitioning recommendations.
In parallel, \textbf{StreamHandler} receives information on active AV sources after requesting it from a component called \textit{AV Registry} via a REST call, and it uses that information to connect to all active AV sources and receive their AV data streams to segment them and store them persistently.
StreamHandler also exposes a REST API that is accessed by SmartViz to request archived AV data from specific sources and points in time. %
The Data Corpus resides at the cloud and is subscribed to all DFB Kafka topics where DatAna publishes SDM-compliant inference results to receive them in real time and archive them internally to make them available for further AI training purposes along with the associated AV data it collects from StreamHandler. The Data Corpus is also subscribed to the DFB Kafka topic that is used by SmartViz to publish user-generated inference result verifications to receive them in real time and update the corresponding archived inference results accordingly. The Data Corpus is also connected to StreamHandler, from which it receives AV data as binary files that are a result of AV stream segmentation.


\subsection{DatAna}

DatAna is a component distributed across all three E2F2C layers, with a separate instance deployed at each infrastructure node.
DatAna is complemented by an MQTT message broker, which is also deployed at each infrastructure node, alongside DatAna.
Each instance of the \href{https://mqtt.org/}{MQTT} message broker is responsible for collecting structured data, i.e., inference results, from the AI components residing on the same layer as the respective MQTT instance.
Specifically, AI components publish their raw inference results to dedicated MQTT topics in real time as they are being produced through the analysis of the AV data streams they receive.
The input inference results of each AI component are formatted as JSON documents according to a dedicated distinct data model that fits the requirements of each AI component. The following are the most notable fields in these data models:
%

 \begin{itemize}[parsep=0em,leftmargin=*,label={--}]
     \item AV source id. The id of the AV source that produced the stream that was analysed to produce the inference result. 
     \item Inference result id. A unique identifier for the inference result.
     \item Timestamps. In case the inference result refers to an instant in time a single timestamp is provided. In case the inference result refers to a period in time, two timestamps are provided, corresponding to the start and end of the time period of the result. All time information is absolute and following the ISO 8601 UTC format. 
 \end{itemize}

 Besides the above, the raw inference results contain other fields that are specific to the needs of each AI component. 

Each DatAna instance residing on the same infrastructure node as an MQTT broker subscribes to the broker’s topics to receive all incoming input AI inference results.
Subsequently, DatAna transforms the input inference results into SDM-compliant counterparts.
Three data models that belong in the collection of smart data models of the SDM standard have been identified to be relevant to MARVEL, which have been modified by adding additional fields to account for the project’s needs:

\begin{itemize}[parsep=0em,leftmargin=*,label={--}]%
    \item \textbf{MediaEvent}: to describe general AI inference results.
    \item \textbf{Alert}: to describe AI inference results that should be perceived as alerts.
    \item \textbf{Anomaly}: to describe AI inference results that should be perceived as detected anomalies.
\end{itemize}
 
DatAna selects autonomously the most appropriate data model to perform the transformation, whose output is then relayed to higher-level layers.

Specifically, the SDM-compliant inference results produced by DatAna at the edge layer are relayed to DatAna at the fog layer and the SDM-compliant inference results produced by DatAna at the fog layer are relayed to DatAna at the cloud layer. The DatAna instance at the cloud layer is responsible for relaying the SDM-compliant inference results it collects from all layers to the DFB by publishing them to the appropriate DFB Kafka topics.

\subsection{DFB}

The DFB resides at the cloud and receives all SDM-compliant inference results published by the DatAna cloud instance and stores them persistently in its ElasticSearch database. The DFB also exposes a REST API to SmartViz to allow it to access all archived inference results in its ElasticSearch database. The DFB receives user-gerenated verifications of inference results from SmartViz when they are published to a dedicated DFB Kafka topic and uses them to update the respective archived inference result entries accordingly. The DFB also accesses a REST API at the HDD for dispatching the currently applied Kafka topic partition information along with associated performance measurements to it. Using the same REST API, the DFB can also receive updated Kafka topic partition allocation that is recommended by the HDD.
SmartViz is subscribed to all DFB Kafka topics where DatAna publishes SDM-compliant inference results to receive them in real time and present them to the user. SmartViz also allows users to verify the inference results they are presented with. SmartViz transmits these user-generated verifications to the DFB by publishing them to a dedicated Kafka topic available at the DFB. 

\subsection{StreamHandler}

StreamHandler resides at the fog and receives AV data streams from all active AV sources (CCTV cameras, network-enabled microphones, AudioAnony and VideoAnony instances) via RTSP. During initialisation, StreamHandler accesses the REST API of the AVRegistry to discover the active AV sources and their details. During operation, StreamHandler consumes the AV RTSP streams and segments them according to a pre-specified time intervals to generate binary documents, suitable for persistent storage. StreamHandler archives the generated AV data files and also exposes a REST API to accept requests from SmartViz about the transmission of AV data from specific AV sources (reference to AV Source id) and from specific points in time. Upon such requests, StreamHandler retrieves the necessary binary files, compiles a unified/edited version of the stream that corresponds to the timeframe requested and generates a link to the said binary file which is to be consumed by SmartViz.

\subsection{HDD}

The HDD exposes a REST API to allow the reception of the currently applied DFB Kafka topic partition information along with associated performance measurements from the DFB. The HDD uses this information as input to calculate an optimised Kafka topic partition allocation and subsequently makes it available to the DFB via its REST API. The exact optimisation method that is implemented by the HDD can be found in \cite{arxiv.2205.09415}.

The DMP has been applied in 5 use cases defined for the needs of the initial version of the MARVEL Integrated framework.

\section{AV Data}\label{sec:avdata}


In the context of the MARVEL framework design activities, certain similarities and overlaps were identified between the functionalities of StreamHandler and those of DFB and DatAna with regards to big data management. However, following an in-depth analysis of the MARVEL framework requirements that the DMP should satisfy, a gap was identified that could not be covered by the DFB and DatAna solutions. This gap was related to the management of audio-visual data. More specifically, the following requirements were established:

\begin{itemize}[parsep=0em,leftmargin=*,label={--}]%
    \item Receive and efficiently archive live streams of audiovisual binary data from all relevant MARVEL sensors, devices and components during system operation.
    \item The persistent storage of archived AV data should comply with high data security standards and data privacy requirements.
    \item Provide access to archived audiovisual binary data to the MARVEL UI (SmartViz) by streaming requested archived audiovisual data upon demand in order to present them to the end-user and in association with relevant inference results produced by MARVEL AI components. 
    \item Support the expansion of the data set of the Data Corpus by relaying selected archived audiovisual data to it. 
\end{itemize}

StreamHandler was found to be in a position to be able to satisfy these requirements and fill the gap by extending its supported data source types, its connectors and data storage capabilities. This course of action was aligned with INTRA’s strategic plan to expand the StreamHandler platform in the direction of audiovisual data management for increased interoperability in order to address additional business cases and reinforce its position in the big data management and smart cities domains. 

\section{Evaluation of the setup}\label{sec:eval}

In this section we summarize the results obtained during the first system integration tests of \MARVEL.

\subsection{DatAna}

%
During the tests performed during the MVP, the performance metrics of a single NiFi instance was measured. Table \ref{tab::datana} summarises the collected measurements for the specified metrics.

\begin{center}
\begin{table}[tbh]
\centering
\caption{Results of the measurements for DatAna component}
\vspace{-0.5em}
\label{tab::datana}
\begin{tabular}{|l|l|}
 \textbf{Metric} & \textbf{Value}  \\ \hline\hline
 Data loss rate & 0  \\  \hline
 Service availability-failed request & 100\% availability\\\hline
 Data access restriction & None\\\hline
 Data throughput & 1.1 MB/s \\\hline
 Response time & 47.1 ms \\\hline
 Number of cluster nodes & 1\\
\end{tabular}
\vspace{-0.5em}
\end{table}
\end{center}

For a more in-depth analysis of performance metrics, there is this Cloudera study \cite{2020cloudera}, which reports how NiFi behaves in terms of scalability and performance (data rates) using very demanding workloads.

\subsection{DFB}

For DFB the following high-level performance indicators were considered:

\begin{itemize}[parsep=0em,leftmargin=*,label={--}]%
    \item \textbf{Data Integrity}: to confirm that advanced encryption mechanisms over end-to-end data transfer will guarantee data integrity. Metric: Data loss rate.
    \item \textbf{Scalability}: to increase the number of modality data streams and verify that performance metrics improve or at least stay the same. Metric: HW speed up.
    \item \textbf{Availability}: to verify that DFB resources are available and discoverable. Metrics: Service availability-failed request, data access restriction.
    \item \textbf{Performance} (for high volume, heterogeneous data streams): to measure different performance metrics under different execution conditions. Metrics: Data transfer latency, data throughput, response time, number of cluster nodes.
\end{itemize}

%
Table \ref{tab::dfb} summarises the collected measurements for the specified metrics.

\begin{center}
\begin{table}[tbh]
\centering
\caption{Results of the measurements for DFB component}
\vspace{-0.5em}
\label{tab::dfb}
\begin{tabular}{|l|l|}
 \textbf{Metric} & \textbf{Value}  \\ \hline\hline
Data loss rate & 0 \\ \hline
HW speed up & - \\ \hline
Service availability-failed request & 100\% availability \\ \hline
Data access restriction & None \\ \hline
Data transfer latency & 5 ms (200 MB/s load) \\ \hline
Data throughput & 605 MB/s \\ \hline
Response time & 5 ms (200 MB/s load) \\ \hline
Number of cluster nodes & 3 \\
\end{tabular}
\vspace{-0.5em}
\end{table}
\end{center}

\subsection{StreamHandler}

Preliminary testing has indicated that StreamHandler is capable of processing at least 3 Full HD AV data streams in parallel with no performance lag when deployed on an infrastructure with 2 CPU cores allocated. 
\subsection{HDD}

\begin{figure}[t!]
    \centering
    \begin{subfigure}[b]{0.49\columnwidth}
        \centering
        \includegraphics[width=\columnwidth]{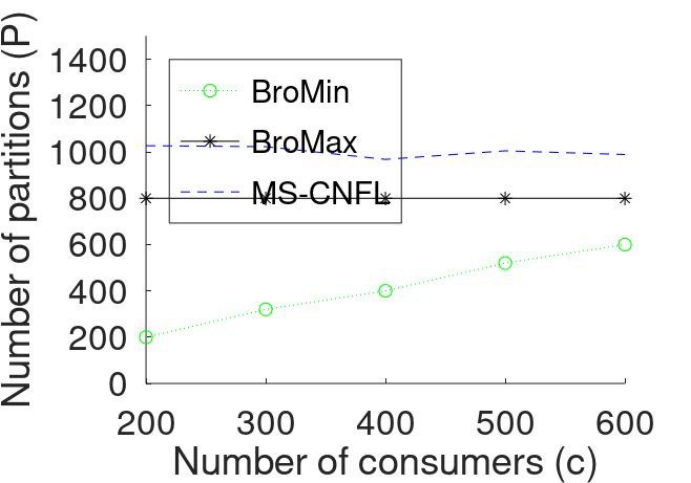}
        \caption{Throughput.}
        \label{fig::hdd-performance}
    \end{subfigure}
    \begin{subfigure}[b]{0.49\columnwidth}
        \centering
        \includegraphics[width=\columnwidth]{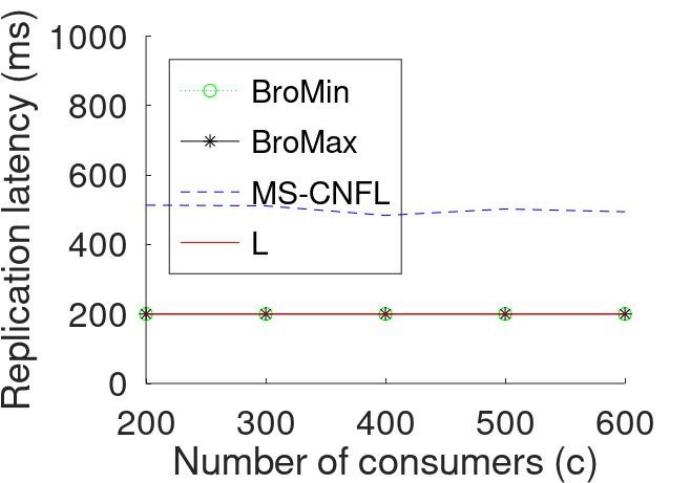}
        \caption{Rep. latency.}
        \label{fig::hdd-latency}
    \end{subfigure}
    \caption{Results of the simulated measurements for HDD component.}
    \label{fig:hdd-performance}
\end{figure}

For the purposes of evaluating the efficiency of  HDD, we took into account the industrial best practices in the related application sectors. We identified Kafka setup guidelines used by credible industrial service providers. For example, Microsoft, recommends that it would be better to constrain the existing partitions per broker (including replicas) to a number not more than $1000$. In another example, Confluent recommends to set the number of partitions per broker to at least $100 \cdot B$. Consequently, combining the essence of these configuration recommendations, we arrive at the following benchmark method, called MS-CNFL: $ P = \min \left( P \in_R [1 ... \frac{1.000 \cdot B}{r} ], P \in_R [1...100 \cdot B] \right)$ and $b \in_R [1 ... B]$, where $\in_R$ denotes uniformly random selection.
%
%
We measure the system throughput, captured by the ultimate number of partitions selected by each algorithm (our algorithms being BroMin and BroMax of \cite{arxiv.2205.09415}), the replication latency, captured the amount of time that is needed to process each message, in the sense of time required for data to be stored or retrieved, the numbers or costs of the application's infrastructure, captured by the number of brokers used in the Apache Kafka cluster, the OS load metric via the open file handles and the unavailability metric via the unavailability time. We perform the measurements for variable number of consumers. Indicatively, we display the performance in terms of throughput (number of partitions) and replication latency, in Fig.~\ref{fig:hdd-performance}. We can see that HDD maintains equivalent numbers of partitions (and therefore throughput), but, at the same time, does not violate the latency constraint (like the benchmark is doing).



\section{Design guidelines} \label{sec::guidelines}

In this Section, we report some useful guidelines which reflect the experiences that we had when building our platform.
\begin{itemize}
\item Distil the data exchange requirements of the involved components to consolidate the necessary I/O interfaces as much as possible and consequently reduce integration complexity.
\item Decouple as much as possible the direct data exchange between pairs of individual component instances to reduce integration complexity, i.e., avoid the use of REST APIs wherever possible and promote the use of pub/sub distributed messaging systems.
\item Implement open, industry-standard approaches for increased interoperability, scalability and expandability.
\item Align the data models used for handling and storing the inference results with the SDM standard in order to improve the visibility and acceptance of the envisioned results.
\item Achieve a versatile, yet consistent  and coherent solution that can support a multitude of different use cases and scenarios and operate on different infrastructure configurations. In our case, this is achieved through the design of the DMP and the specification of an adaptive reference ``AI Inference Pipeline'' architecture. The DMP is fully scalable and interoperable as it can be adapted to incorporate virtually any number of edge and fog nodes, while it can handle data emerging from any MARVEL component (e.g., anonymisation components, AI components) at any layer of the E2F2C continuum.
\item Handle multimodal raw (AV) and structured (inference results) data by collecting from and distriburing among multiple endpoints both in real time and asynchronusly via persistent storage mechanisms.
\item Maintain an up-to-date comprehensive documentation of the specifications for all implemented I/O interfaces and data models using version control. In our case, a GitLab repository was used for this purpose.
\end{itemize}


%
%
%

\balance

\bibliographystyle{IEEEtran}
\bibliography{samplebibfile}

\end{document}